\documentclass[twocolumn]{aastex62}
\usepackage{natbib}
\usepackage[version=4]{mhchem}
\graphicspath{{./}{figures/}}

\received{2020 August 31}
\revised{2020 March 27}
\accepted{2020 April 11}
\published{2020 May 20}

\shorttitle{Exo-oceanography}
\shortauthors{Olson et al.}

\begin{document}

\title{Oceanographic Considerations for Exoplanet Life Detection}

\correspondingauthor{Stephanie L. Olson}
\email{stephanieolson@uchicago.edu}

\author{Stephanie L. Olson}
\affiliation{Department of the Geophysical Sciences, University of Chicago}

\author{Malte Jansen}
\affiliation{Department of the Geophysical Sciences, University of Chicago}

\author{Dorian S. Abbot}
\affiliation{Department of the Geophysical Sciences, University of Chicago}

\begin{abstract}

Liquid water oceans are at the center of our search for life on exoplanets because water is a strict requirement for life as we know it. However, oceans are dynamic habitats---and some oceans may be better hosts for life than others. In Earth's ocean, circulation transports essential nutrients such as phosphate and is a first-order control on the distribution and productivity of life. Of particular importance is upward flow from the dark depths of the ocean in response to wind-driven divergence in surface layers. This `upwelling' returns essential nutrients that tend to accumulate at depth via sinking of organic particulates back to the sunlit regions where photosynthetic life thrives. Ocean dynamics are likely to impose constraints on the activity and atmospheric expression of photosynthetic life in exo-oceans as well, but we lack an understanding of how ocean dynamics may differ on other planets. We address this issue by exploring the sensitivity of ocean dynamics to a suite of planetary parameters using ROCKE-3D, a fully coupled ocean-atmosphere GCM. Our results suggest that planets that rotate slower and have higher surface pressure than Earth may be the most attractive targets for remote life detection because upwelling is enhanced under these conditions, resulting in greater nutrient supply to the surface biosphere. Seasonal deepening of the mixed layer on high obliquity planets may also enhance nutrient replenishment from depth into the surface mixed layer. Efficient nutrient recycling favors greater biological activity, more biosignature production, and thus more detectable life. More generally, our results demonstrate the importance of considering oceanographic phenomena for exoplanet life detection and motivate future interdisciplinary contributions to the emerging field of exo-oceanography.

\end{abstract}


\section{Introduction} \label{sec:intro}
Water is an essential ingredient for life as we know it \citep{mckay_requirements_2014}. For this reason, the potential existence of a liquid water ocean defines the Habitable Zone concept that guides our search for life in the Universe \citep{kasting_habitable_1993}. However, oceans are dynamic habitats---and oceanographic processes have additional and far-reaching implications for habitability that remain largely unexplored. 
\begin{figure*}[ht!]
\centering
\includegraphics[width=\textwidth]{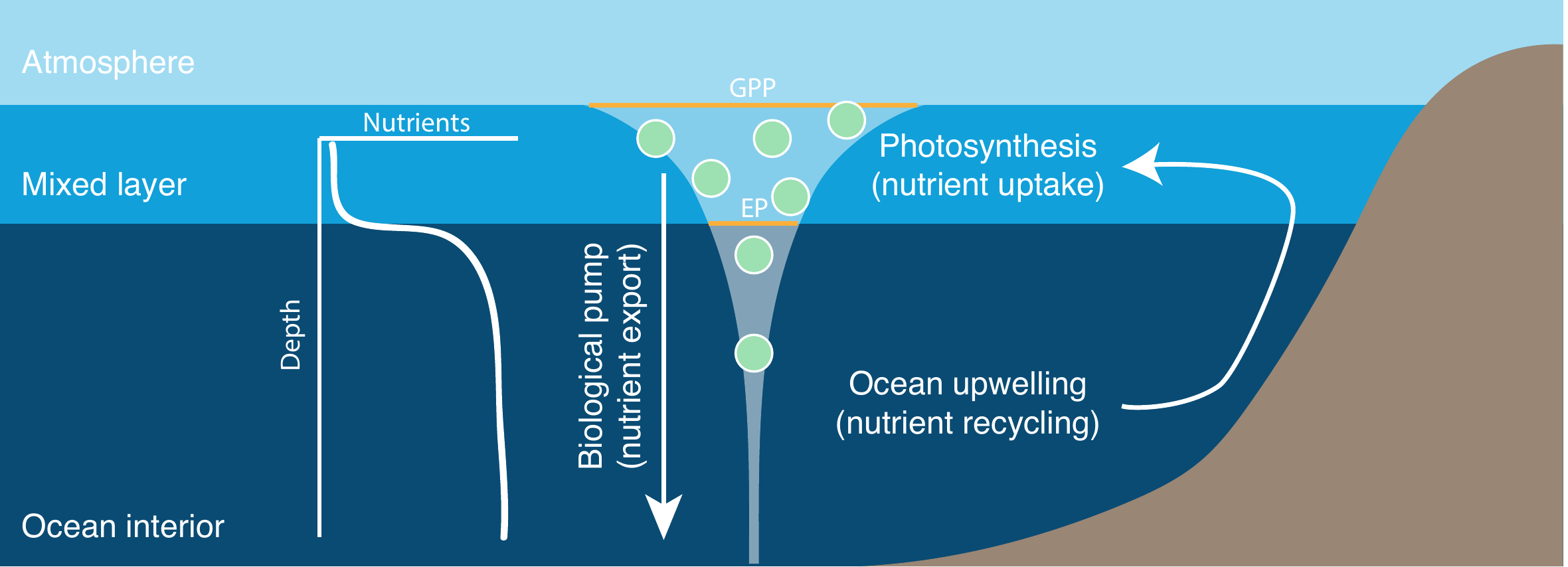}
\caption{\textbf{Schematic of ocean nutrient cycling.} Nutrients such as phosphate are consumed by photosynthetic life in the sunlit portion of the ocean and are gravitationally exported to depth through the settling of particulate organics, resulting in surface waters that are nutrient-depleted and deep waters that are nutrient-rich. The fraction of total biomass production (Gross Primary Productivity; GPP) that settles out of the mixed layer is referred to as export production (EP). EP allows the isolation of photosynthetic O$_2$ and reduced organic carbon, which is essential for both the surface accumulation of O$_2$ and the production of reduced biogenic gases like CH$_4$ at depth in Earth's ocean---but EP also removes nutrients from the surface environment and necessitates that nutrients are recycled via ocean upwelling to sustain the biosphere (see \citealt{libes_introduction_2009}).}
\label{fig:nutrients}
\end{figure*}
Recent studies have investigated the importance of considering ocean heat transport for regulating climate and elucidating the boundaries of the Habitable Zone \citep{hu_role_2014, cullum_importance_2014, cullum_importance_2016, yang_ocean_2019}, but the significance of ocean circulation is not limited to climate influences. 

Ocean circulation is also a primary control on the distribution of biological activity at Earth's surface. Briefly, life in Earth's ocean is concentrated in the shallow sunlit portion of the water column where photosynthesis is viable. The chemical reaction corresponding to photosynthesis can be represented as:  \begin{equation} \ce{CO2 + H2O ->[$hv$] CH2O + O2} \end{equation} where CH$_2$O is a simple representation of biomass. In reality, biomass is chemically complex and includes a number of additional bioessential elements (nutrients), including N and P, and it has a C:N:P ratio of 106:16:1 on average today \citep{redfield_biological_1958}. The availability of essential nutrients thus limits the amount of photosynthesis that can occur. The majority of photosynthetic biomass is degraded by respiration \begin{equation} \ce{CH2O + O2 -> CO2 + H2O} \end{equation} in the shallow ocean but a small fraction escapes degradation by settling through the water column, bringing the nutrients consumed during photosynthesis with it. This export of organic particulates from the shallow ocean, referred to as the `biological pump' \citep{sundquist_ocean_1985, meyer_influence_2016}, preserves the chemical disequilibrium produced from stellar energy during photosynthesis by physically separating reduced organic carbon from photosynthetic O$_2$. Separation of reduced C and photosynthetic O$_2$ stimulates a diversity of microbial metabolisms within the ocean interior and marine sediments, including CH$_4$ production by methanogens \citep{canfield_pathways_1993, reeburgh_oceanic_2007, libes_introduction_2009}. Export production is thus essential for the oxygenation of our atmosphere and the net production of other putative biosignature gases such as CH$_4$ on Earth \citep{logan_terminal_1995}, but efficient removal of biomass from the sunlit portion of the ocean requires a mechanism for replenishing nutrients lost to depth. The primary mechanism for nutrient replenishment to the mixed layer of Earth's ocean is upward flow of deeper water to the surface ocean (upwelling). 

Upwelling is primarily a wind-driven phenomenon that occurs in regions where the horizontal ocean current diverges. Conservation of mass requires upwelling of water from below in response to this divergence. For example, upwelling occurs where winds drive ocean currents off the coast of a continent that obstructs lateral flow. Upwelling also occurs at low latitudes as the consequence of opposing directions of Coriolis deflection on either side of the equator. Vertical mixing of the ocean is otherwise disfavored because the ocean is stably stratified with respect to density, with warm, less dense water on top of cold, denser water. A critical impact of upwelling is that it brings nutrient-rich water up to the surface from the deep ocean. As a result, photosynthetic life is overwhelmingly concentrated in upwelling regions of Earth's ocean today \citep{behrenfeld_photosynthetic_1997}, and biological activity is directly modulated by surface winds \citep{rykaczewski_influence_2008}. This cycle of nutrient uptake in the shallow ocean, export to depth, and recycling via upwelling is summarized in Figure {\ref{fig:nutrients}}. 

The importance of surface winds is not limited to their role in large-scale ocean circulation patterns. The winds also influence global biogeochemical cycles through their impact on the mixed layer depth. The mixed layer is the portion of the water column that is homogenized by turbulence and is in direct contact with the overlying atmosphere. The depth of Earth's mixed layer varies spatially, but its volume is a small fraction of the present-day global ocean volume (a few percent). Dramatic deepening of the mixed layer reduces the average light levels a photosynthetic cell experiences in its lifetime and upon death may increase its exposure time to photosynthetic or photochemically derived oxidants that favor its decomposition. In sum, very deep mixed layers may reduce gross primary productivity (GPP) via light inhibition as well as export production (EP) via enhanced recycling internal to the mixed layer \citep{sverdrup_conditions_1953, li_mechanistic_2017}, ultimately limiting net production of biosignature gases like O$_2$ and CH$4$ that depend on the physical separation of photosynthetic oxidants and reduced organic matter. A  mixed layer that is shallow compared to light penetration depths may thus favor remotely detectable biospheres by enhancing productivity and export---but, ironically, efficient export reinforces the critical importance of ocean upwelling for sustaining biospheric productivity by returning nutrients to the surface.

Although life on other planets is likely to differ from life on Earth, photosynthetic life will require nutrients for the construction of its biomolecules regardless of the details of its biochemistry. Moreover, it is likely that these nutrients would tend to gravitationally accumulate at depth in exo-oceans. It is thus reasonable to expect that ocean circulation patterns may be a first-order control on the activity of photosynthetic life on inhabited exoplanets as well. These relationships have practical implications for the detectability of life elsewhere because the most active surface biospheres with the greatest export fractions will have the greatest potential to influence the spectral appearance of their host planets and will thus be the most detectable biospheres \citep{schwieterman_exoplanet_2018, krissansen-totton_disequilibrium_2018}. Conversely, subsurface life, low productivity biospheres, and/or biospheres in which biosignatures are either accumulated at depth or efficiently recycled within the ocean will be very challenging to detect because biosignature production and communication to the atmosphere will be limited under these circumstances.   

A productive biosphere is an insufficient prerequisite for detectability because biogenic gases within the ocean will not be recognizable with telescopes. Remotely detectable marine biospheres also require the transport of biogenic gases from the ocean environment to the atmosphere via sea-air gas exchange. The global sea-to-air flux of O$_2$ is described by: \begin{equation} F_{O_{2}}= k_{O_{2}}A([O_{2}] - [O_{2}]_{sat}), \end{equation} where $A$ is the surface area of the ocean, $[O_{2}] - [O_{2}]_{sat}$ reflects oceanic O$_2$ super- or under-saturation with respect to the overlying atmosphere, and $k_{O_{2}}$ is the O$_2$ gas exchange constant. $k_{O_{2}}$ is sensitive to wind stress, sea surface temperature, and the extent of sea ice cover. If the exchange flux of O$_2$ is small compared to biological fluxes within the ocean and/or its destruction within the atmosphere, disequilibrium between the ocean and the atmosphere can be maintained \citep{kasting_box_1991, olson_quantifying_2013} with potentially important ramifications for remote life detection, including the possibility of `false negatives' for life despite large-scale biological O$_2$ production \citep{reinhard_false_2017}. Similarly, extensive biological production of CH$_4$ in the ocean does not necessarily manifest as high levels of atmospheric CH$_4$ because biological CH$_4$ oxidation internal to the ocean may severely limit its flux to the atmosphere \citep{reeburgh_oceanic_2007, beal_high_2011, olson_limited_2016, reinhard_oceanic_2020}, depending on oxidant availability, ocean upwelling rates, and the areal extent of sea ice.

Despite their importance, we lack a rigorous understanding of how ocean upwelling, the mixed layer depth, and the transfer of marine biosignatures to the atmosphere may differ among the diversity of habitable exoplanets. In other words, we do not know which planetary scenarios are most conducive to the development of remotely detectable oceanic biospheres---or whether these scenarios are observationally distinguishable. Placing constraints on exo-ocean circulation patterns would aid in identifying the most favorable targets for detailed characterization. This knowledge would also provide useful context for evaluating the vulnerability of a particular planet to a biosignature false negative and assist in assigning significance to inherently ambiguous non-detections \citep{reinhard_false_2017}. 

Whereas detecting exo-oceans will be feasible with future instruments \citep{robinson_detecting_2010, lustig-yaeger_detecting_2018}, directly characterizing ocean dynamics and marine habitats will not be possible. It is thus necessary to understand the sensitivity of ocean circulation patterns to observable planetary parameters---and to understand the uncertainty introduced by other factors that may be difficult to constrain remotely. As a first step, we use a general circulation model (GCM) to quantify the sensitivity of global upwelling and other biogeochemically significant oceanographic quantities to a broad range of planetary parameters (Sections \ref{sec:period}--\ref{sec:salinity}). We then discuss how these oceanographic constraints may affect biospheric productivity and the detectability of life on inhabited planets differing from our own (Section \ref{sec:bio_disc}). We conclude by offering recommendations regarding the most favorable targets for exoplanet life detection as well as discussing observational prospects for assessing the likelihood of a false negative vs. a true negative in the face of an ambiguous non-detection (Section \ref{sec:obs_disc}).

\section{Model description} \label{sec:methods}
We perform our calculations using ROCKE-3D \citep{way_resolving_2017}, a fully coupled ocean-atmosphere GCM that is modified from the NASA Goddard Institute for Space Studies (GISS) ModelE2 \citep{schmidt_configuration_2014}. Of particular relevance for this study, ROCKE-3D includes a thermodynamic-dynamic sea ice model and the versatile SOCRATES radiative transfer scheme \citep{edwards_efficient_1996, edwards_new_2007}. See \citet{way_resolving_2017} for a detailed description of ROCKE-3D and its parent model. The model is publicly available from the NASA GISS ModelE repository. 

Our ROCKE-3D simulations use 4${^{\circ}}$x5${^{\circ}}$ latitude-longitude resolution with 40 vertical layers in the atmosphere (up to 0.1 mbar) and 10 depth layers in the ocean (down to 1360 m). Ocean eddies are unresolvable at this resolution. Eddy fluxes are parameterized following the Gent-McWilliams-Redi skew flux scheme \citep{redi_oceanic_1982, gent_isopycnal_1990, gent_parameterizing_1995, griffies_gentmcwilliams_1998}.  

We spun up each model scenario to a steady-state, which we diagnosed by the achievement of a global radiative balance of 0 $\pm$ 0.2 W m$^{-2}$ averaged over the last 10 years of the run. We further confirmed steady-state by checking for stable temperature and salinity in the abyssal ocean. Radiative balance was typically achievable in 500 model years, but reaching steady-state required modestly longer run times for some model scenarios. All of the data we show are averaged over the last 10 years of each simulation independent of the total run time.   

\subsection{Baseline Configuration} \label{sec:baseline} 
Our `baseline planet' configuration resembles present-day Earth in many ways (see Table \ref{table:stdconfig}). We adopt Earth values for the mass, radius, and surface gravity of our baseline planet. Additionally, our baseline planet has a 24-hour rotation period and orbits a sun-like star with a 365-day period. Our baseline planet receives an Earth-like stellar irradiation of 1360 W m$^{-2}$, but we assume that the planet's obliquity and eccentricity are both zero to eliminate complications arising from seasonally variable irradiation. This choice ultimately reduces run times and allows a greater exploration of parameter space. 

Like Earth, our baseline planet has a surface pressure of 1 atm at sea level. Unlike Earth, however, our baseline planet lacks O$_2$ (and O$_3$) and instead has an N$_2$ atmosphere (\textgreater99\%) with trace (pre-industrial) levels of CO$_2$ and CH$_4$. The combination of zero obliquity and modest CO$_2$ yields a climate that is somewhat cooler than that of present-day Earth, particularly at high latitudes, but is nonetheless Earth-like and habitable (Table \ref{table:atmdata}). 

\begin{figure*}[p]
\centering
\includegraphics[width=\textwidth]{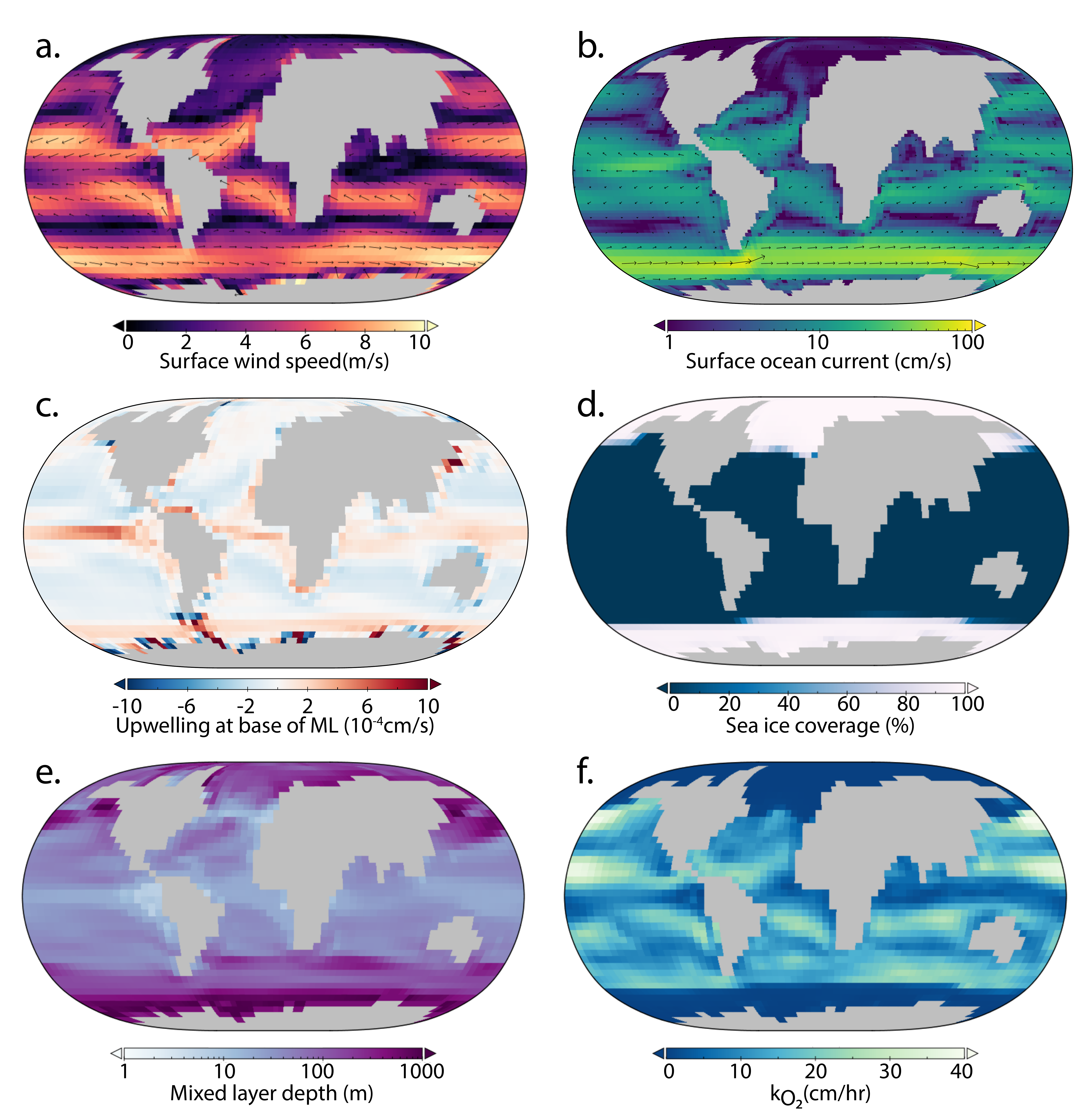}
\caption{\textbf{Key oceanographic parameters for the Earth-like baseline planet.} Shown are: surface winds (a), surface ocean velocities (b), vertical velocities (upwelling) at the base of the mixed layer (c), sea ice coverage (d), mixed layer depth (e), and the oxygen exchange coefficient (f).}
\label{figure:baseline}
\end{figure*}

The distribution of land masses and ocean bathymetry on our baseline planet is based on present-day Earth with a few exceptions (Figure \ref{figure:baseline}). Most notably, we have implemented a `bathtub' ocean bathymetry \citep{way_climates_2018}. This ocean has deepened shelves (591 m) compared to our ocean and it has a flat bottom that is shallower than Earth's ocean (1360 m). Moreover, this ocean bathymetry eliminates several small and/or shallow seas such as the Mediterranean, Baltic, Black Sea, Red Sea, and Hudson Bay by designating these areas as landmass. Unlike \citet{way_climates_2018}, we have also eliminated Baffin Bay by removing the island of Greenland. In combination, these changes to the continental configuration and ocean bathymetry allow examination of a greater diversity of habitable climates by avoiding numerical stability issues that can arise on icy planets with shallow oceans \citep{way_climates_2018}. 

\begin{deluxetable}{lll}[t]
\caption{Baseline planet parameters}
\label{table:stdconfig}
\tablehead{
\colhead{\bf{Parameter}} & \colhead{\bf{Baseline}}}
\startdata
Rotation period  		 & 24 hours 	   \\
Orbital period   		 & 365 days 	   \\
Mass			 		 & $M_{\Earth}$	   \\
Radius 		    		 & $r_{\Earth}$	   \\
Surface gravity 		 & 9.8 m/s$^{2}$   \\
Surface pressure 		 & 1 atm    	   \\
Obliquity        		 & 0$^{\circ}$     \\
Eccentricity 	 		 & 0$^{\circ}$	   \\
Stellar spectrum 	     & Sun        	   \\
Stellar irradiation 	 & 1360 W m$^{-2}$ \\
Map 	         		 & Modified Earth  \\
Ocean depth				 & 1360 m		   \\
Salinity         		 & 35 PSU          \\
\enddata
\end{deluxetable}

\begin{figure*}[t]
\centering
\includegraphics[width=0.85\textwidth]{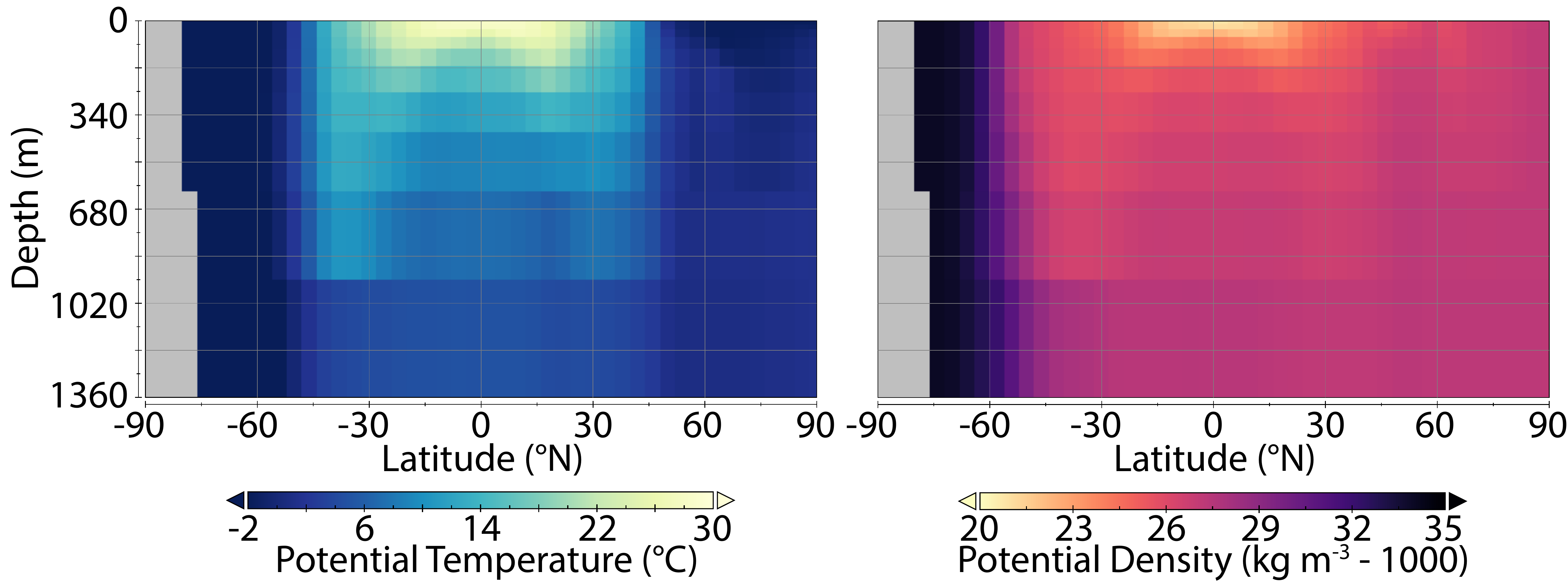}
\caption{\textbf{Temperature (left) and Density (right) structure of the baseline ocean.} Warm, low density water generally sits atop colder, denser water. This stable stratification breaks down at high latitude, allowing particularly dense water to sink. Vertical density gradients in the ocean are ultimately a reflection of lateral density gradients because the deep ocean is filled with the densest water from the surface.}
\label{figure:thermo}
\end{figure*}

\begin{deluxetable*}{lrrr}[pb]
\caption{\textbf{Climate data with relevance for planetary  habitability.} Multipliers in experiment descriptions are with respect to  our baseline (Earth-like) planetary scenario (Table \ref{table:stdconfig}). Equator-to-pole temperature contrast is calculated as the average of the two latitude bands straddling the equator minus the average of the two poles. Note that snow/ice cover here is inclusive of continental snow and ice in addition to sea ice, but some plots include only sea ice.}
\label{table:atmdata}
\tablehead{\colhead{\bf{Experiment}} & \colhead{\bf{Ave. Temp (C)}} & \colhead{\bf{Eq-Pole $\Delta$T (C) }} & \colhead{\bf{Snow/Ice Cover (\%)}}}
\startdata
\textit{\textbf{Baseline}} & 10.04 & 80.26 & 20.27  \\
\hline
\textit{\textbf{Rotation Rate}} \\
0.1x  & 10.25 & 53.7  & 4.11  \\
0.5x  & 10.72 & 72.5  & 16.0  \\
2x    & 3.11  & 108.5 & 32.24 \\
\hline
\textit{\textbf{Surface pressure}} \\ 		 
0.5x & -16.97 & 109.1 & 50.18 \\
2x   & 20.21  & 51.46 & 4.79  \\
5x 	 & 6.00   & 66.02 & 19.13 \\
10x  & 2.56   & 58.73 & 20.09 \\
\hline
\textit{\textbf{Orbital Obliquity}} \\	
15$^{\circ}$  & 10.82 & 71.96 & 18.8 \\
30$^{\circ}$  & 16.85 & 53.58 & 5.90 \\
45$^{\circ}$  & 19.33 & 10.08 & 4.29 \\
\hline
\textit{\textbf{Stellar Irradiation}} \\	
0.74x   & -46.79 & 80.00 & 89.76 \\
0.92x   & -13.54 & 93.80 & 46.82 \\
1.1x   & 27.30  & 52.96 & 2.69 \\
\hline
\textit{\textbf{Ocean Salinity}} \\	
0.1x & 6.92 & 82.06 & 26.18 \\
0.5x & 8.13 & 83.15 & 23.68 \\
2x  & 15.74 & 55.53 & 6.96 \\
\hline
\textit{\textbf{Planet Radius}} \\	
1.5x & 4.67 & 109.93 & 29.59 \\
2x 	 & 6.22 & 127.13 & 27.07 \\
\hline
\enddata
\end{deluxetable*}

\subsection{Baseline Circulation} \label{sec:baseline} 
Before exploring ocean sensitivity to various planetary and oceanic parameters, it is useful to briefly summarize the salient features of Earth's ocean circulation and highlight each in our baseline model scenario. Our discussion is deliberately simplified and qualitative; see \cite{marshall_atmosphere_2008} or \cite{vallis_atmospheric_2017} for a more thorough discussion of Earth's ocean and atmospheric circulation.

The large-scale atmospheric circulation is driven by an unequal distribution of stellar energy between the equator and the poles, and it is strongly modulated by planetary rotation (the `Coriolis effect'). In combination, these phenomena manifest in surface winds with a distinct pattern of reversals with increasing latitude (Figure \ref{figure:baseline}a). At low latitudes in each hemisphere, surface winds consistently blow from east to west; these easterly winds are referred to as the `trade winds.' Westerly winds prevail in the mid-latitudes, and the high latitudes experience easterly winds at the surface.

Ocean circulation is strongly influenced by these surface winds. However, wind-driven surface currents in the ocean do not simply mirror the winds in either speed or direction (Figure \ref{figure:baseline}b). There are two reasons. First, ocean currents experience additional Coriolis deflection with respect to wind stress such that bulk wind-driven `Ekman transport' in the upper ocean is perpendicular to the wind stress at the surface. The easterly component of the tropical trade winds in each hemisphere therefore yields equatorial divergence within the surface ocean despite equatorial convergence in the atmosphere at the surface. Continents also obstruct oceanic flow. The combination of these barriers and rotational effects leads to subtropical ocean gyres with subcircular motion \citep{enderton_explorations_2009}. These gyres are associated with subduction of surface water and nutrient poor conditions \citep{rodgers_extratropical_2003}. Nutrient replenishment via  upwelling, as discussed above, is concentrated in regions where the winds drive divergent surface flows. These regions are primarily along the coasts of continents and along the equator (Figure \ref{figure:baseline}c).

In addition to the wind-driven circulation in the upper ocean, the transport of dissolved gases and nutrients in the ocean is affected by the deep ocean overturning circulation. Water cools as it moves poleward, weakening the density stratification of the ocean (Figure \ref{figure:thermo}) and resulting in deeper wind mixed layers (Figure \ref{figure:baseline}). Exclusion of salt when sea ice forms at high latitude can also increase the density of seawater---and this cold, salty water tends to sink. Sinking of dense surface water to the abyssal ocean is necessarily balanced by upwelling elsewhere. In Earth's present-day ocean much of this upwelling occurs in the Southern ocean, again driven by the winds, with additional upwelling distributed primarily over regions with strong turbulent mixing (e.g., \citealt{marshall_closure_2012, wunsch_vertical_2004}).

\begin{deluxetable}{lll}[tb]
\caption{Sensitivity Experiments}
\label{table:exps}
\tablehead{
\colhead{\bf{Parameter}} & \colhead{\bf{Minimum}} & \colhead{\bf{Maximum}}}
\startdata
Rotation period   		 & 12 hours       		  & 10 days        \\
Radius 			 		 & $r_{\Earth}$	   		  & 2$r_{\Earth}$  \\
Surface pressure  		 & 0.5 atm   	   		  & 10 atm  	   \\
Stellar irradiation		 & 1000 w/m$^{2}$ 		  &	1500 w/m$^{2}$ \\
Obliquity		  		 & 0$^{\circ}$ 	   		  & 45$^{\circ}$   \\
Salinity	 	  		 & 3.5 PSU		    	  & 70 PSU	       \\
\enddata
\end{deluxetable}

\subsection{Sensitivity Experiments} \label{sec:exp}
We examine the sensitivity of this baseline ocean circulation to: radius, surface pressure, rotation rate, obliquity, stellar irradiation, and ocean salinity. We change each parameter from our baseline experiment in isolation, with the exception of a few parameters that we co-vary. We outline our procedures and underlying assumptions for these experiments below, and Table \ref{table:exps} summarizes the ranges for each parameter.  

We vary planet radius up to 2x Earth's radius ($r_{\Earth}$). This is a narrow range compared to the radii of known exoplanets, but it is generously inclusive of the radii of planets that are potentially rocky and Earth-like \citep{rogers_most_2015}. Upon changing radius, we also update planet mass, surface gravity, and surface pressure. Following the empirical relationship derived by \citet{kopparapu_habitable_2014}, we assume that planetary mass is related to its radius by: 
\begin{equation} 
	\bigg( \frac{M_{p}}{M_{b}} \bigg) = 0.968 \bigg( \frac{r_{p}}{r_{b}} \bigg)^{3.2} 
\end{equation} 
where $M_b$ and $r_b$ are the mass and radius of our baseline planet ($M_b$ and $r_b$ are equal to $M_{\Earth}$ and $r_{\Earth}$, respectively).

Surface gravity is in turn related to both the planetary mass and radius by: 
\begin{equation} 
	\bigg( \frac{g_{p}}{g_{b}} \bigg)= \bigg( \frac{M_{p}}{M_{b}} \bigg)\bigg( \frac{r_{b}}{r_{p}} \bigg)^{2} 
    \end{equation} 
where $g_b$ refers to the surface gravity on our baseline planet, $g_{\Earth}$ (9.8 m s$^{-2}$). 

Surface pressure is proportional to surface gravity, and it is further modulated by the surface area of the planet ($A_p$) and the mass of the overlying atmosphere ($m_p$): 
\begin{equation} 
	\bigg( \frac{P_{p}}{P_{b}} \bigg)= \bigg( \frac{m_{p}}{m_{b}} \bigg) \bigg( \frac{g_{p}}{g_{b}} \bigg) \bigg( \frac{A_{b}}{A_{p}} \bigg) 
\end{equation} 
where $m_b$ and $A_b$ represent the surface area and atmospheric mass of our baseline planet. We scale the mass of the atmosphere as the surface area evolves with radius such that $m_{p}/A_{p}$ = $m_{b}/A_{b}$. Substitution yields:
\begin{equation} 
	\bigg( \frac{P_{p}}{P_{b}} \bigg)=  0.968 \bigg( \frac{r_{p}}{r_{b}} \bigg)^{1.2}
\end{equation}

We note that this formulation diverges somewhat from that of \citet{kopparapu_habitable_2014} because they assumed that $m_p$ is proportional to $M_p$. Their scaling between planetary and atmospheric mass may be a reasonable approximation, but we instead opt to preserve $m/A$ for each of our radius experiments and modify atmospheric mass in isolation in subsequent sensitivity analyses. We did this by changing surface pressure between 0.5 and 10 atm for constant surface gravity, mass, and radius. We further assumed a fixed recipe for air (i.e., we kept gas mixing ratios constant rather than adjusting $p$N$_2$ in isolation with constant abundances of trace greenhouse gases). 

\subsection{Oceanographic Metrics}
Our analysis focuses on 10-year, global averages of several oceanographic properties of biogeochemical significance: 
\begin{enumerate} 
\item \textbf{wind stress.} We calculate global average windstress from model output as \begin{equation} \label{eq:stress} \tau = C_{D} \rho_{atm} U^{2}, \end{equation} where $\rho_{atm}$ is atmospheric density, which increases proportional to surface pressure for atmospheres of constant composition. $U$ is surface wind speed (m/s). We assume that $C_{D}$, the wind drag coefficient, is constant across the planetary parameter space we explore. We exclude land cells from our wind stress calculation but we do not account for the effects of sea ice, which modulates the transfer of wind stress to the underlying ocean in ROCKE-3D \citep{zhang_modeling_2000}.  
\item \textbf{density stratification.} We leverage the surface-to-deep potential density contrast, $\Delta \sigma$, as a proxy for the stability of the density stratification. We simply calculate $\Delta \sigma$ as the average potential density of the surface ocean layer minus the global-average potential density of the bottom ocean layer. Potential density is the density that a parcel of water would have if adiabatically brought to the surface; whereas in situ density varies with depth (pressure) in the ocean, potential density is not a function of depth and simply reflects differences in temperature and salinity. The vertical potential density contrast is ultimately a reflection of horizontal equator-to-pole density gradients in the surface ocean because the deep ocean is filled with the densest waters from the surface that sink to the deep ocean as part of the global overturning circulation.
\item \textbf{the depth of the mixed layer.} The depth of the mixed layer in ROCKE-3D varies in space and time, and is calculated using the K profile parameterization (KPP) scheme \citep{large_oceanic_1994}. 
\item \textbf{ocean upwelling.} Upwelling is presented as globally summed upwelling at the base of the mixed layer. Although we spatially average the mixed layer depth, our upwelling sum accounts for spatial variability in the depth of the mixed layer and is calculated as the area-weighted sum of upward flow (cm$^{3}$ s$^{-1}$) in the depth layer containing the base of the mixed layer for each latitude and longitude position. Summing upwelling at fixed depth yields similar results. Upwelling is classified as equatorial if it occurs in the two latitude bands of grid cells straddling the equator (\textless 4$^{\circ}$ N/S) and upwelling is classified as coastal if any of the eight adjacent cells is land. Cells may be counted as both equatorial and coastal, but are only counted once toward the global total. 
\item \textbf{sea-air gas exchange constant}. We calculate the gas exchange constant for O$_2$ from model output following \citet{wanninkhof_relationship_2014} with minor modifications to account for ice cover and variable surface pressure: \begin{equation} \label{eq:kO2} k_{O_{2}} = 0.251(1-f_{ice}) U^{2} \bigg( \frac{P_{p}}{P_{b}} \bigg)  \bigg( \frac{Sc_{O_{2}}}{660} \bigg) ^{-0.5} \end{equation} where $f_{ice}$ is the fractional ice cover. The Schmidt number, $Sc_{O_{2}}$, is equal to 568 at 20 $^{\circ}$C and is described by a fourth-order polynomial with respect to sea surface temperature  \citep{wanninkhof_relationship_2014}.
\end{enumerate} 

\section{Results} \label{sec:results}

\begin{figure*}[t]
\centering
\includegraphics[width=\textwidth]{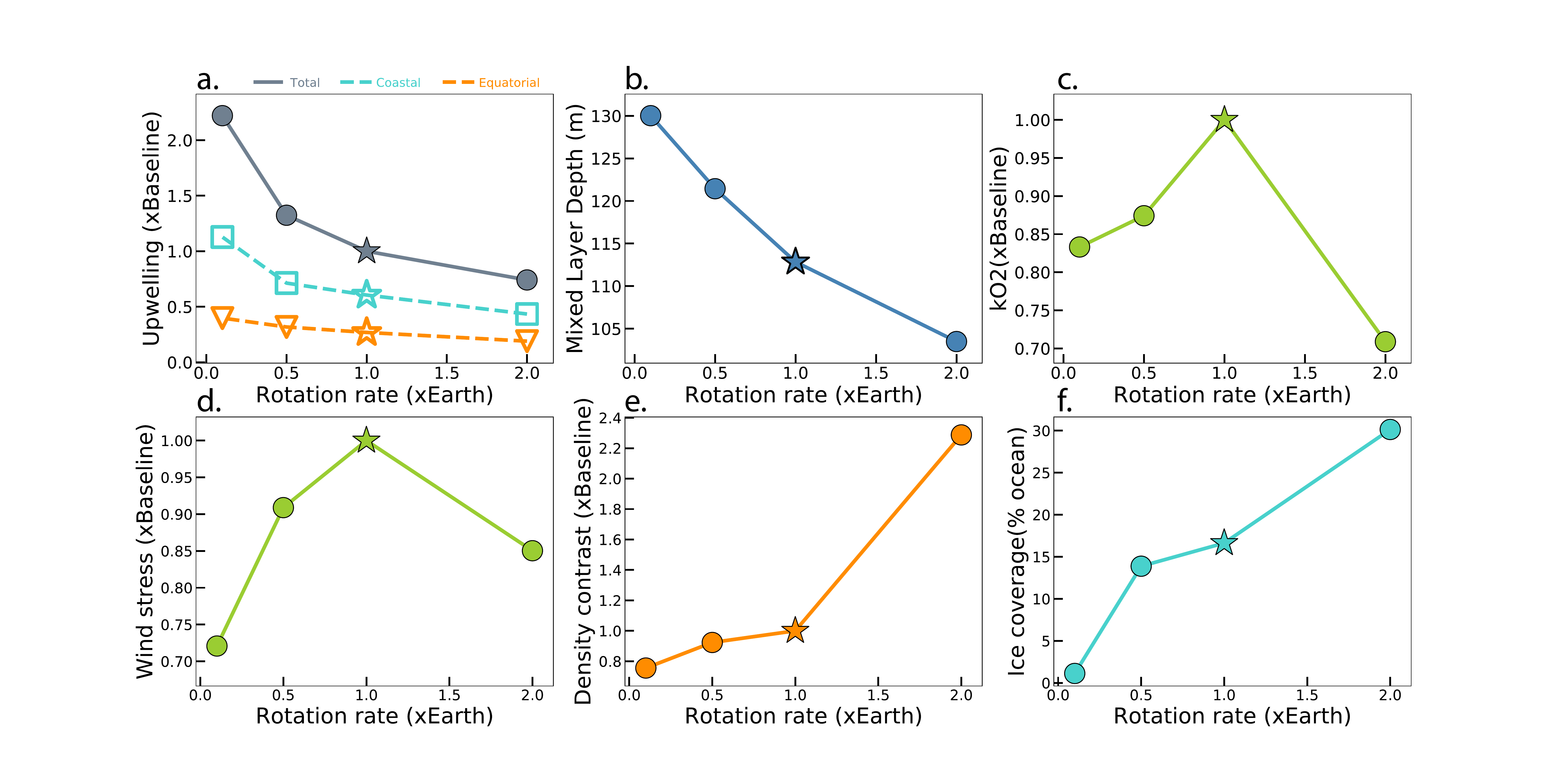} 
\caption{\textbf{Ocean-atmosphere sensitivity to rotation rate}, including: globally summed upwelling at the base of the mixed layer (a), global-average mixed layer depth (b), global-average oxygen gas exchange constant (c), global-average wind stress over ocean cells (d), global-average surface-to-deep density contrast (e), and global sea ice coverage (f). In each panel, the star denotes the Earth-like baseline planet. Upwelling, $k_{O_2}$, wind stress, and the density contrast are normalized to their baseline values for ease of comparison. In (a), the filled grey circles represent the global total, the open blue squares are the coastal upwelling contribution to that total, and the open orange triangles are the equatorial upwelling component. All data are averaged over the last decade of the simulations.}
\label{fig:rote}
\end{figure*}

\begin{figure*}[thbp]
\centering
\includegraphics[width=\textwidth]{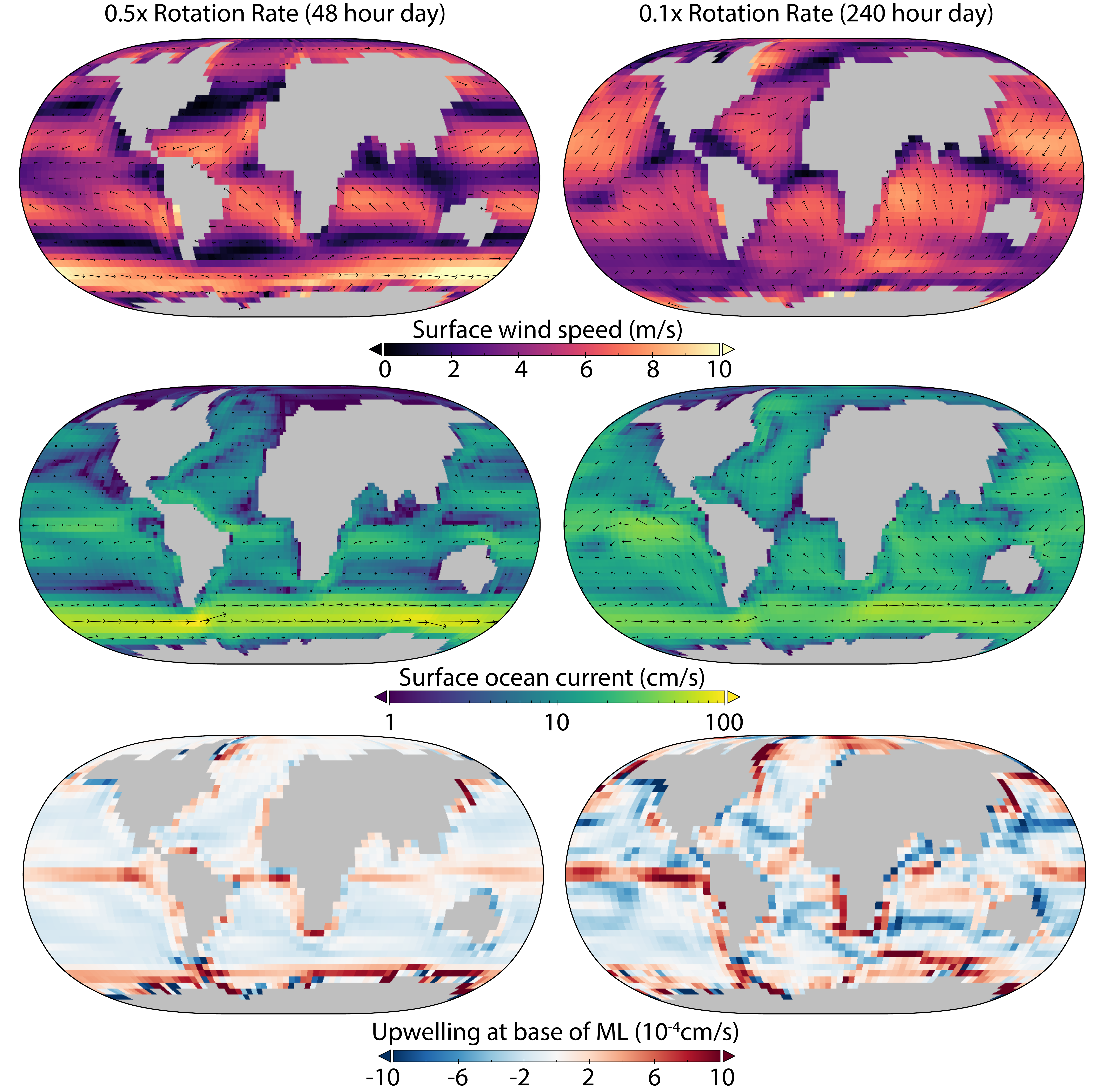}
\caption{\textbf{Surface currents (top) and upwelling (bottom) for a 48 hour day and a 240 hour day scenarios.} Weakening Coriolis results in a restructuring of atmospheric circulation \citep{kaspi_atmospheric_2015, komacek_atmospheric_2019} with major consequences for wind-driven ocean circulation.}
\label{figure:circulation}
\end{figure*}

\begin{figure*}[t]
\centering
\includegraphics[width=\textwidth]{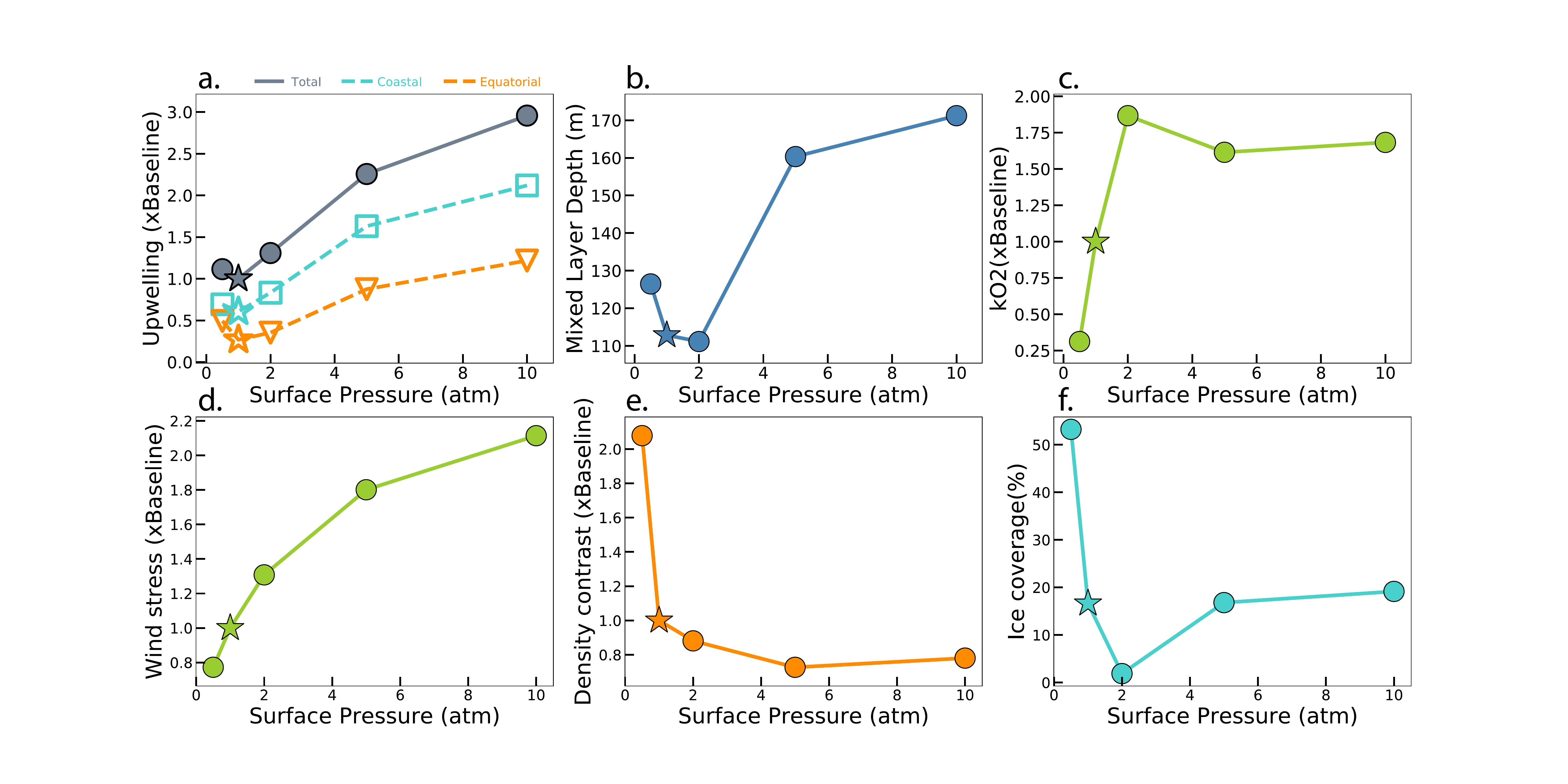} \label{fig:press}
\caption{\textbf{Ocean-atmosphere sensitivity to surface pressure}, including: globally summed upwelling at the base of the mixed layer (a), global-average mixed layer depth (b), global-average oxygen gas exchange constant (c), global-average wind stress over ocean cells (d), global-average surface-to-deep density contrast (e), and global sea ice coverage (f). In each panel, the star denotes the Earth-like baseline planet. Upwelling, $k_{O_2}$, wind stress, and the density contrast are normalized to their baseline values for ease of comparison. In (a), the filled grey circles represent the global total, the open blue squares are the coastal upwelling contribution to that total, and the open orange triangles are the equatorial upwelling component. All data are averaged over the last decade of the simulations.}
\label{fig:press}
\end{figure*}

\begin{figure*}[t]
\centering
\includegraphics[width=\textwidth]{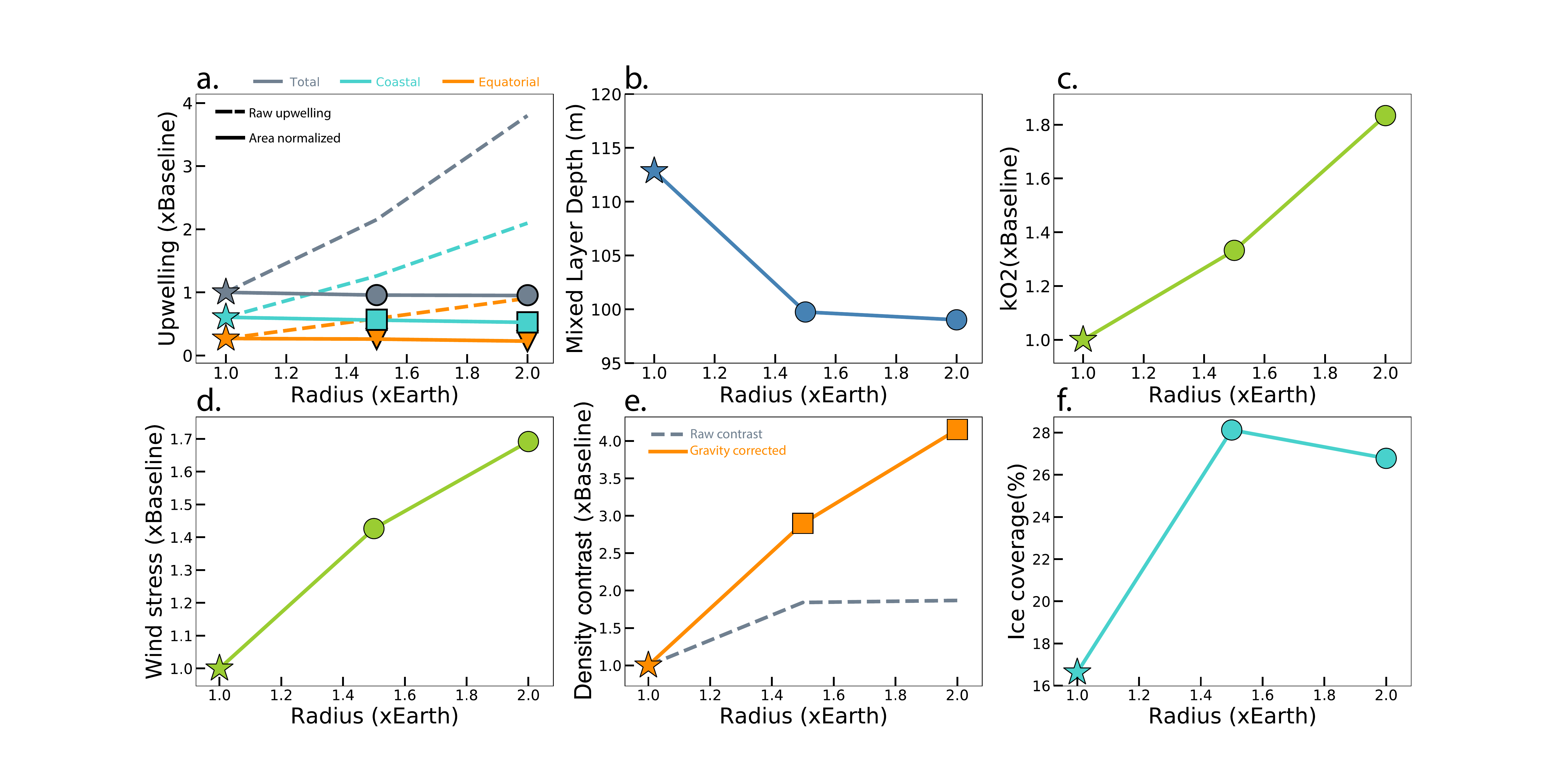}
\caption{\textbf{Ocean-atmosphere sensitivity to planet radius}, including: globally summed upwelling at the base of the mixed layer (a), global-average mixed layer depth (b), global-average oxygen gas exchange constant (c), global-average wind stress over ocean cells (d), global-average surface-to-deep density contrast (e), and global sea ice coverage (f). In each panel, the star denotes the Earth-like baseline planet. Upwelling, $k_{O_2}$, wind stress, and the density contrast are normalized to their baseline values for ease of comparison. In (a), the filled grey circles represent the global total, the filled blue squares are the coastal upwelling contribution to that total, and the filled orange triangles are the equatorial upwelling component. The open symbols share the same color and symbol associations with total, coastal, and equatorial upwelling, but these data have been normalized to surface area, which increases as $r^{2}$. In (e), filled circles represent our simple stratification metric, $\Delta\sigma$, as elsewhere in this text, and the open squares have been corrected for gravity influences on buoyancy as planetary radius is increased ($g\Delta\sigma$). All data are averaged over the last decade of the simulations.}
\label{fig:radius}
\end{figure*}

\begin{figure*}[t]
\centering
\includegraphics[width=\textwidth]{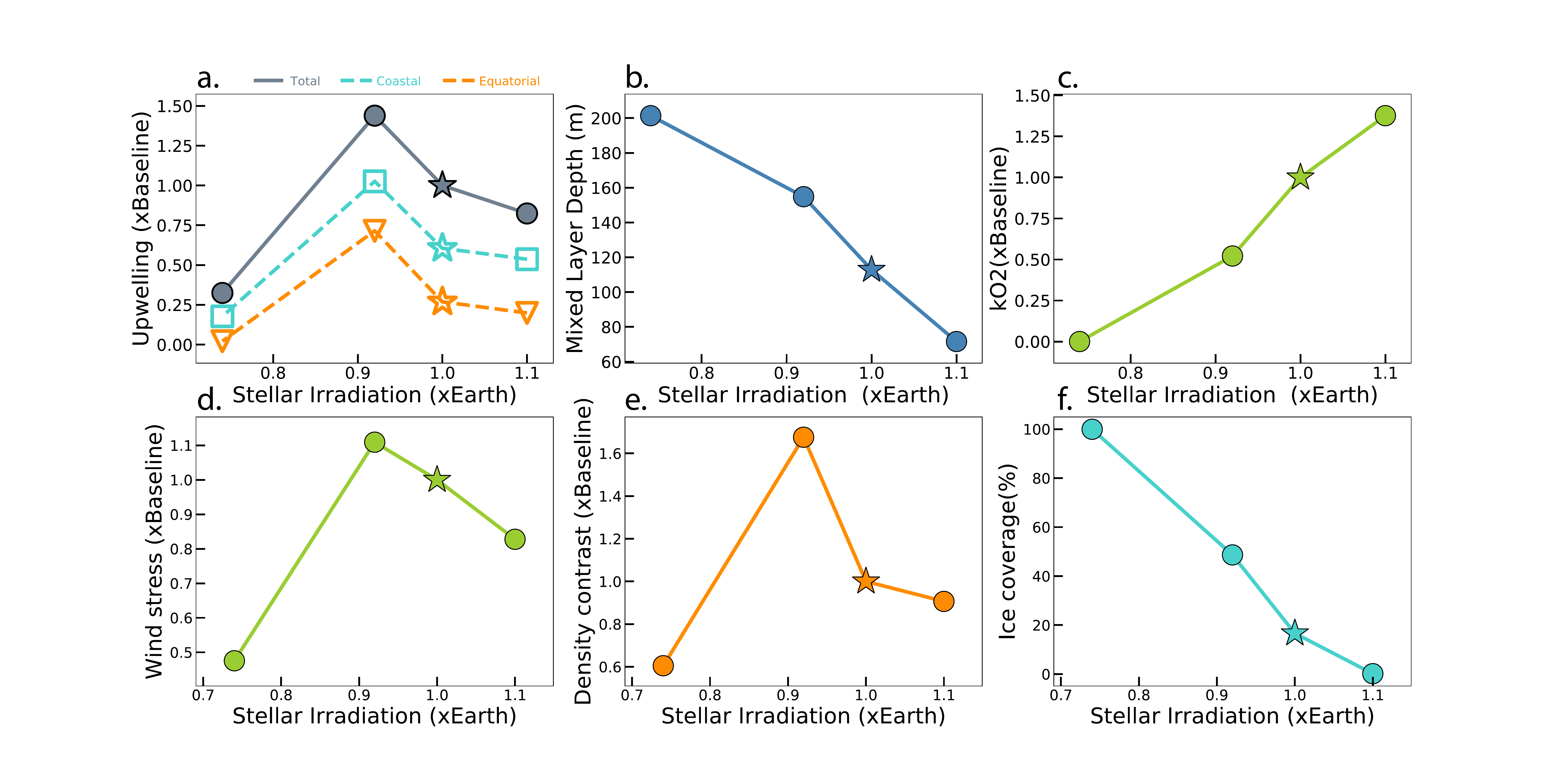}
\caption{\textbf{Ocean-atmosphere sensitivity to stellar irradiation}, including: globally summed upwelling at the base of the mixed layer (a), global-average mixed layer depth (b), global-average oxygen gas exchange constant (c), global-average wind stress over ocean cells (d), global-average surface-to-deep density contrast (e), and global sea ice coverage (f). In each panel, the star denotes the Earth-like baseline planet. Upwelling, $k_{O_2}$, wind stress, and the density contrast are normalized to their baseline values for ease of comparison. In (a), the filled grey circles represent the global total, the open blue squares are the coastal upwelling contribution to that total, and the open orange triangles are the equatorial upwelling component. All data are averaged over the last decade of the simulations.}
\label{fig:insol}
\end{figure*}

\begin{figure*}[p]
\centering
\includegraphics[width=\textwidth]{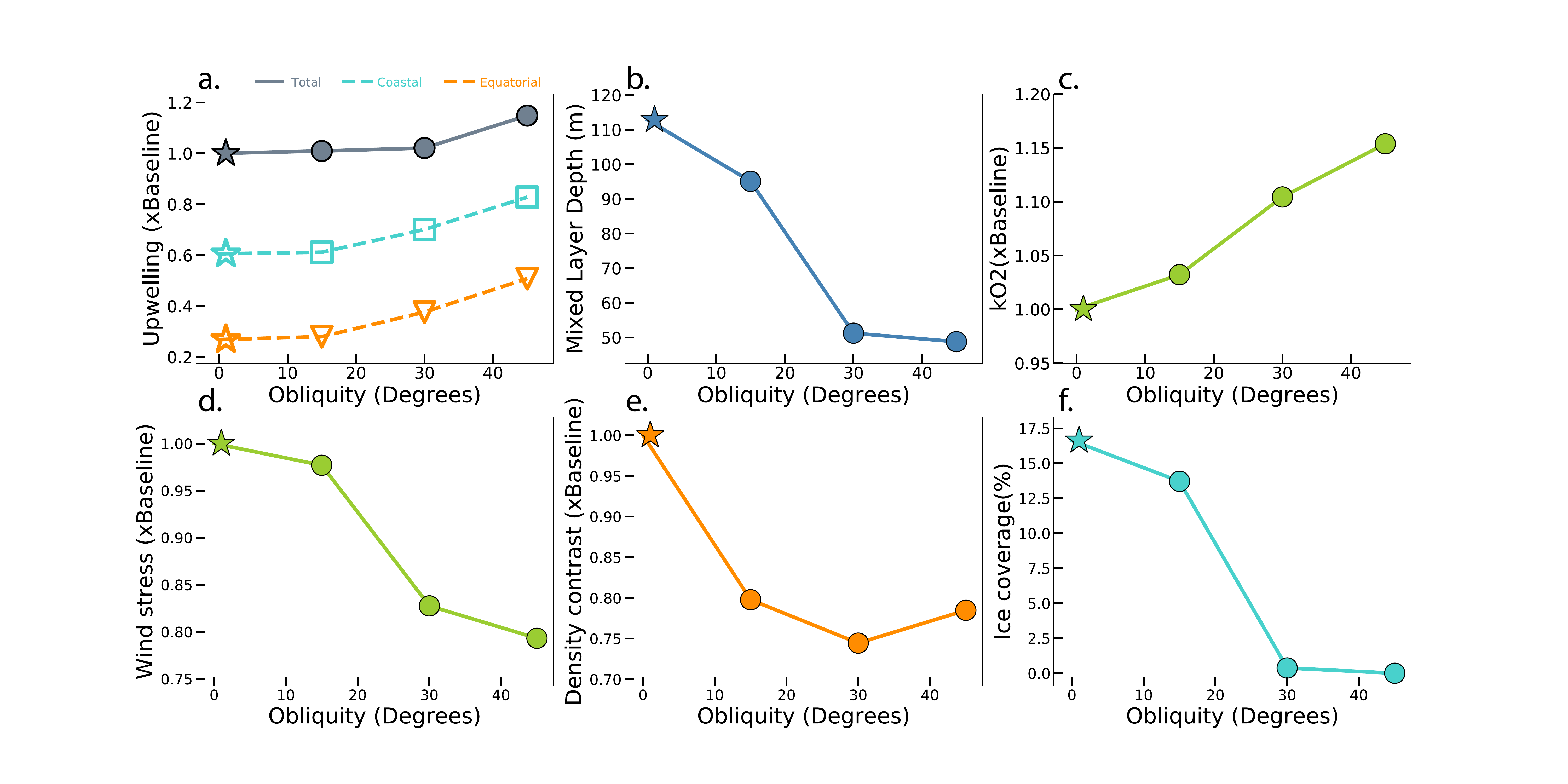}
\caption{\textbf{Ocean-atmosphere sensitivity to orbital obliquity}, including: globally summed upwelling at the base of the mixed layer (a), global-average mixed layer depth (b), global-average oxygen gas exchange constant (c), global-average wind stress over ocean cells (d), global-average surface-to-deep density contrast (e), and global sea ice coverage (f). In each panel, the star denotes the baseline planet, which is generally Earth-like except that it has zero obliquity. Upwelling, $k_{O_2}$, wind stress, and the density contrast are normalized to their baseline values for ease of comparison. In (a), the filled grey circles represent the global total, the open blue squares are the coastal upwelling contribution to that total, and the open orange triangles are the equatorial upwelling component. All data are averaged over the last decade of the simulations.}
\label{fig:obliq}
\end{figure*}

\begin{figure*}[h]
\centering
\includegraphics[width=\textwidth]{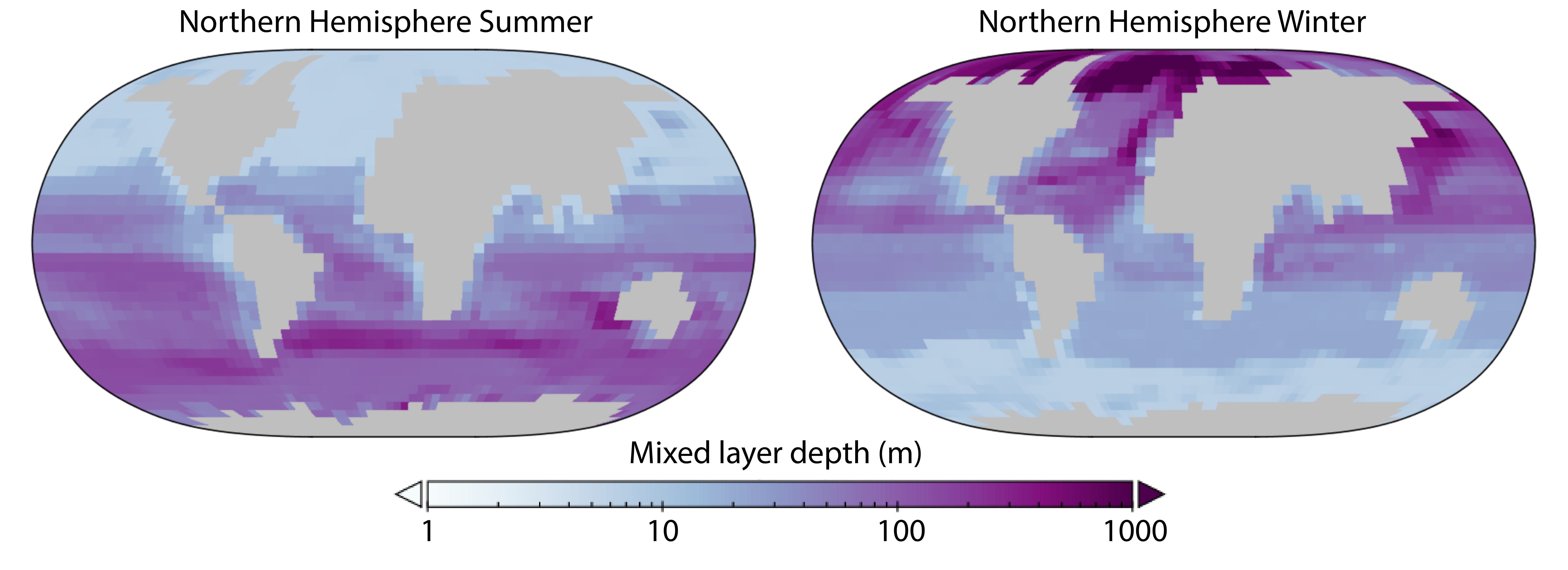}
\caption{\textbf{Seasonality in the mixed layer depth on a planet with 45$^{\circ}$ obliquity.} The mixed layer depth locally varies by \textgreater 2 orders of magnitude across the year. This seasonal deepening of the mixed layer allows direct entrainment of nutrients from depth independent of upwelling. }
\label{fig:seasons}
\end{figure*}

\begin{figure*}[t]
\centering
\includegraphics[width=\textwidth]{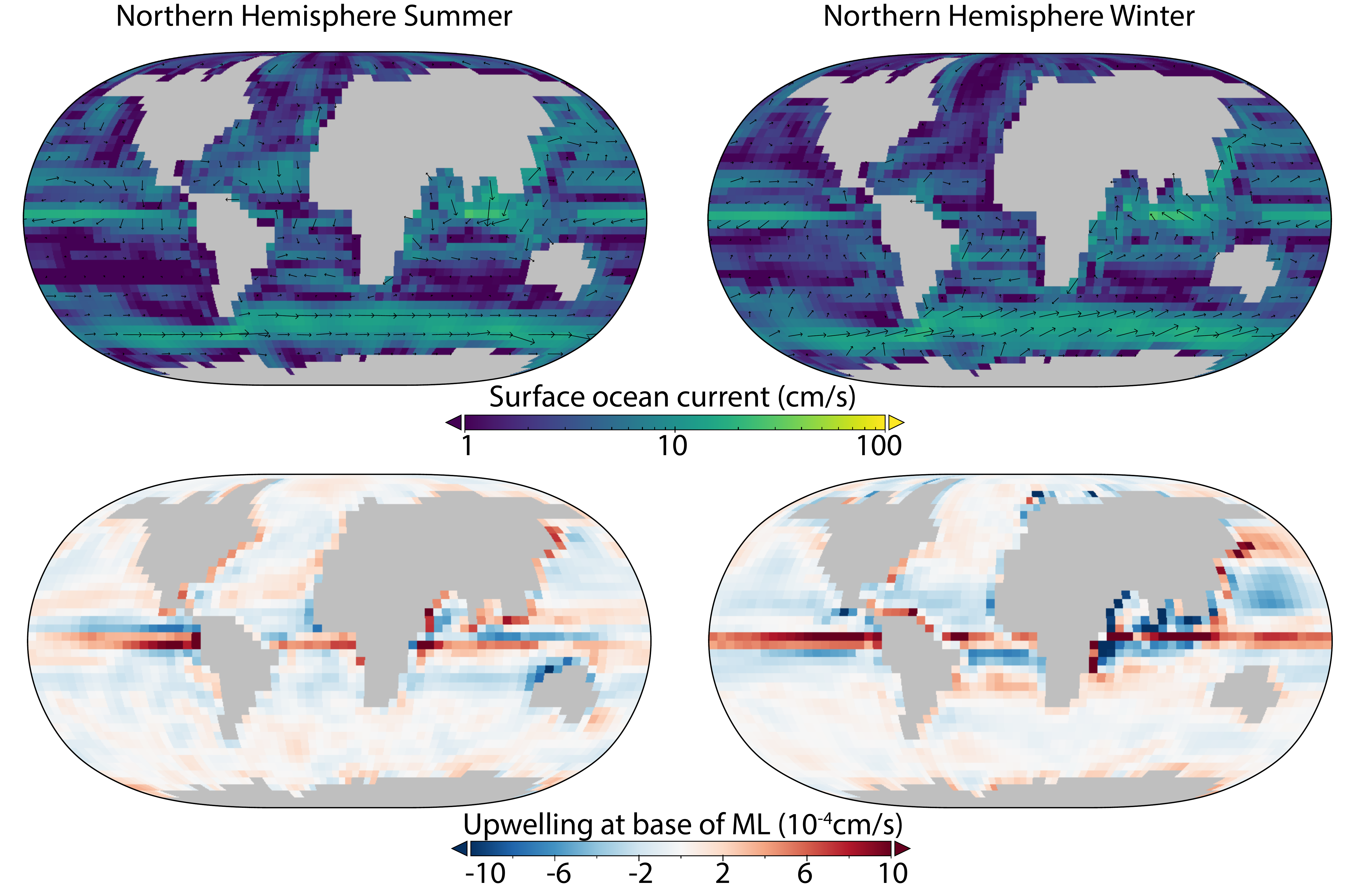}
\caption{\textbf{Seasonality in surface currents (top) and upwelling (bottom) on a planet with 45$^{\circ}$ obliquity.} Some surface currents reverse directional seasonally because the summer pole becomes warmer than the equator. The spatial distribution of upwelling shifts as a result.}
\label{fig:o45_circ}
\end{figure*}

\begin{figure*}[t]
\centering
\includegraphics[width=\textwidth]{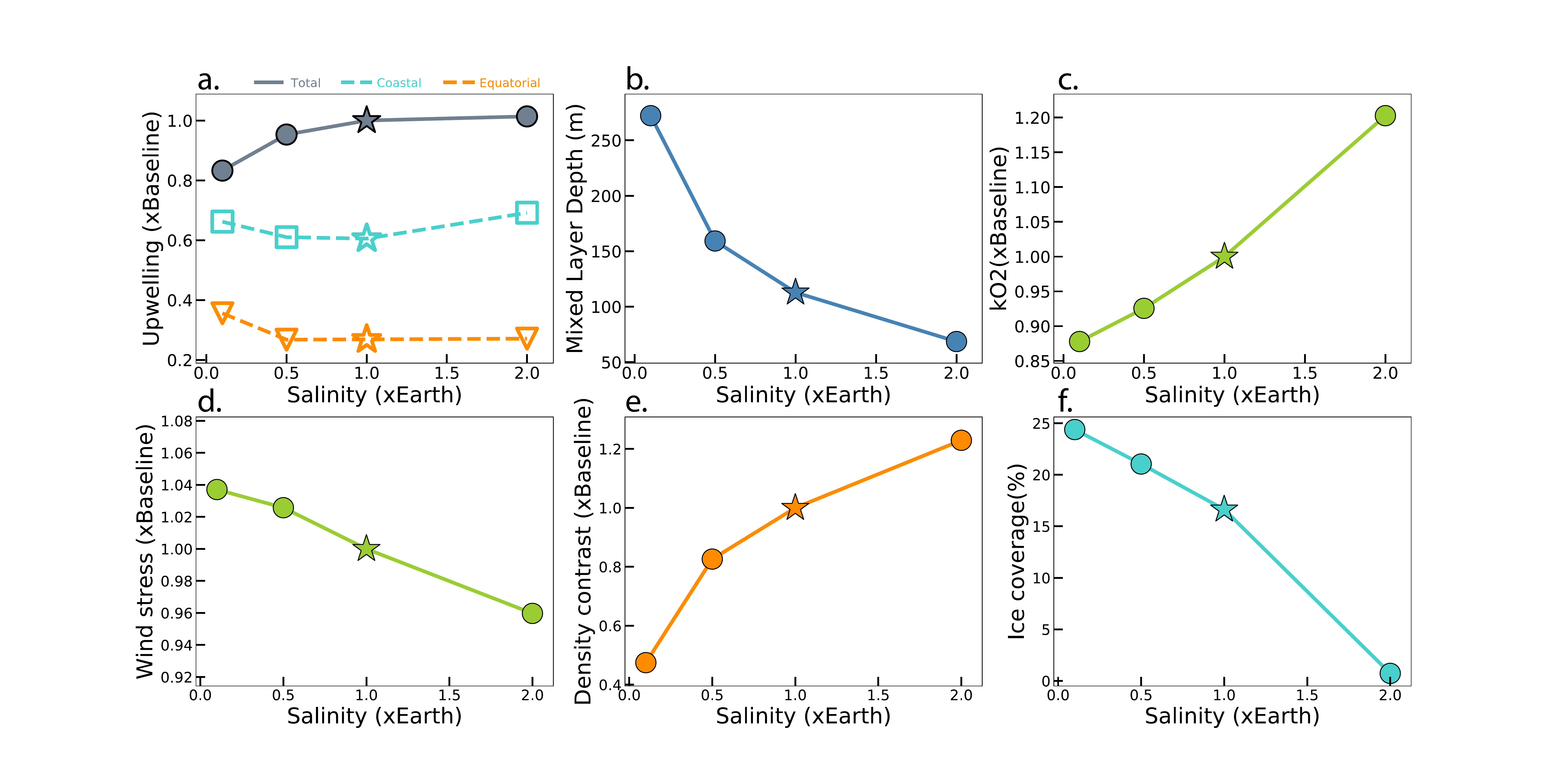}
\caption{\textbf{Ocean-atmosphere sensitivity to ocean salinity}, including: globally summed upwelling at the base of the mixed layer (a), global-average mixed layer depth (b), global-average oxygen gas exchange constant (c), global-average wind stress over ocean cells (d), global-average surface-to-deep density contrast (e), and global sea ice coverage (f). In each panel, the star denotes the Earth-like baseline planet. Upwelling, $k_{O_2}$, wind stress, and the density contrast are normalized to their baseline values for ease of comparison. In (a), the filled grey circles represent the global total, the open blue squares are the coastal upwelling contribution to that total, and the open orange triangles are the equatorial upwelling component. All data are averaged over the last decade the simulations.}
\label{fig:salt}
\end{figure*}

\subsection{Rotation Rate} \label{sec:period}
Globally integrated upwelling at the base of the mixed layer increases with decreasing rotation rate (equivalently, increasing day length; Figure \ref{fig:rote}a). For modest changes in rotation rate, changes in upwelling are qualitatively predicted by the expected response of the wind-driven surface ocean Ekman transport, which is described by: 
\begin{equation} \label{eq:eckman}
    V = \frac{\tau}{\rho f}
\end{equation}
where $V$ is the magnitude of the horizontal wind-driven transport integrated over the ocean surface boundary layer and $f$ is the Coriolis parameter. The Coriolis parameter is defined as: 
\begin{equation} 
	f = 2 \Omega sin(\varphi)
\end{equation} 
where $\Omega$ represents the planet's rotation rate and $\varphi$ is latitude. On global average, $f$ is simply equal to the rotation rate. Upwelling is primarily driven by divergence of this wind-driven Ekman transport and should therefore be inversely proportional to rotation rate for a given wind stress. 

Our simulations span multiple atmospheric circulation regimes including the familiar Earth-like circulation regime and a slow rotation regime characterized by weak Coriolis influences  \citep{kaspi_atmospheric_2015, komacek_atmospheric_2019}. The resulting changes in the surface winds, together with increasing deviations from Ekman balance (which only holds for relatively rapidly rotating planets), leads to significant differences in the spatial distribution of upwelling in the ocean. Although global upwelling uniformly increases with decreasing rotation rate, Coriolis deflection of Ekman transport no longer sustains equatorial divergence in our simulation with a 240-hour day. Instead of equatorial upwelling, this simulation produces convergence and downwelling over most of the equatorial Pacific (Figure \ref{figure:circulation}).  

Rotation rate also influences the globally averaged mixed layer depth. The mixed layer depth modestly increases with decreasing rotation rate in our experiments despite slower winds and decreasing wind stress at slow rotation rates (Figure \ref{fig:rote}b,d). This counter-intuitive result appears to arise due to enhanced atmospheric meridional heat transport \citep{kaspi_atmospheric_2015, komacek_atmospheric_2019}. The result is a smaller equator-to-pole temperature gradient as rotation rate decreases (Table \ref{table:atmdata}), which in turn leads to a weaker density stratification at low latitudes because the density of deep water is set by the density of surface seawater at high latitudes where deep water is formed.

The global-average coefficient for O$_2$ exchange with the atmosphere does not respond monotonically to increasing rotation rate. $k_{O_{2}}$ increases with increasing rotation rate from 0.1--1x Earth's rotation rate as wind stress increases, but the increase is partially compensated for by cooling and expanding ice. With further increases in rotation rate, the combination of decreasing wind stress and increasing sea ice results in a sharp reduction of sea-air gas exchange. 

\subsection{Surface Pressure} \label{sec:pressure}
The depth of the mixed layer and global upwelling at the base of the mixed layer both increase with increasing surface pressure beyond 1 atm (Figure \ref{fig:press}a). This relationship primarily arises from increased wind stress with increasing surface pressure, which allows the winds to exert greater influence on ocean dynamics via Equation \ref{eq:eckman}. Wind stress is also strongly sensitive to surface wind speed (Equation \ref{eq:stress}). Wind speed decreases with increasing surface pressure as the consequence of friction, but these changes in wind speed are smaller than the changes in atmospheric density in our experiments. The pressure effect thus dominates the wind stress response (Figure \ref{fig:press}d). Greater wind stress contributes to enhanced wind-driven ocean circulation, including surface divergence. The result is more upwelling beneath higher density atmospheres. Deviation from this trend at low surface pressure likely arises due to a large increase in sea ice cover (Figure \ref{fig:press}f).

Increasing surface pressure initially increases surface temperatures due to the combined effects of higher greenhouse gas abundances and pressure broadening, but Rayleigh scattering eventually yields cooling for surface pressure above 2 atm \citep{keles_effect_2018, komacek_atmospheric_2019}. Meanwhile, meridional atmospheric heat transport increases with increasing surface pressure, resulting in a smaller difference between equatorial and polar temperatures \citep{kaspi_atmospheric_2015}. This reduced latitudinal temperature contrast mutes the vertical density stratification of the ocean on average because the deep ocean is ultimately filled with the densest waters that sink from the surface and fill the deep ocean (Figure \ref{fig:press}e). 

Sea-air gas exchange is initially favored by increasing surface pressure due to the combined effects of increasing wind stress, warming, and reductions in sea ice cover (Equation \ref{eq:kO2}). At surface pressures much higher than 1 atm, cooling ultimately leads to an expansion of sea ice cover, which limits ocean-atmosphere connectivity.

We note that the surface pressure at which climatic trends reverse will be sensitive to atmospheric composition \citep{komacek_atmospheric_2019}, and the details of the relationship between surface pressure, climate, and ocean dynamics may differ on planets with differing greenhouse gas abundances.

\subsection{Radius} \label{sec:radius} 
Unlike our other sensitivity analyses, we did not vary planetary mass and radius in isolation; instead, we co-varied mass, gravity, and surface pressure as we changed radius (see discussion in Section \ref{sec:exp}). 

We found that global upwelling increases with increasing radius (Figure \ref{fig:radius}a, dashed lines). However, we note that this trend is eliminated when global upwelling is normalized to surface area which increases as $r^{2}$ (Figure \ref{fig:radius}a, solid lines). In other words, upwelling per unit area is nearly constant despite an absolute increase in the global sum on larger planets. 

The global-average mixed layer depth decreases slightly with increasing radius despite an increase in wind stress (Figure \ref{fig:radius}b,d). There are two potential reasons. First, the equator-to-pole temperature contrast increases with increasing planetary radius (\cite{kaspi_atmospheric_2015}; Table \ref{table:atmdata}), increasing the potential for strong vertical temperature contrast. Moreover, the dynamically relevant buoyancy stratification is also enhanced directly when surface gravity is increased (Figure \ref{fig:radius}e). Although we have simplistically adopted the density contrast, $\Delta\sigma$, as a stratification metric for global comparisons between simulations, the dynamically relevant metric is the buoyancy stratification, which is proportional to $g\Delta\sigma$. The open squares in Figure \ref{fig:radius}e show $g\Delta\sigma$ and reflect the gravitational influence of changing planetary mass and radius on buoyancy stratification. This effect strongly stabilizes stratification in opposition to the effect of increased wind stress on the mixed layer depth.  

\subsection{Stellar Irradiation} \label{sec:insolation}
Varying stellar irradiation from 1000 to 1500 W m$^{-2}$ assuming constant $p$CO$_2$ yields climates that range from snowball states to ice-free states. Global upwelling at the base of the mixed layer increases with decreasing stellar irradition---but upwelling drops off as sea ice cover increases to 100\% (Figure \ref{fig:insol}a,f). These changes in upwelling generally mirror changes in globally averaged wind stress (Figure \ref{fig:insol}d), with variable modulation by ice cover that is not accounted for in our wind stress metric. We also note that ROCKE-3D neglects geothermal heat input at the bottom of the ocean, which may be an important influence on ocean dynamics on ice-covered worlds \citep{ashkenazy_dynamics_2013, jansen_turbulent_2016}. 

The mixed layer gets shallower with increasing stellar irradiation above the snowball threshold (Figure \ref{fig:insol}b). This trend is opposite to the relationship between the mixed layer and warming on global average in some of our other experiments. The reason for this difference is that warming induced by increasing surface pressure or reducing rotation rate enhances meridional heat transport and tends to decrease the equator-to-pole temperature contrast (Table \ref{table:atmdata}). These effects generally weaken ocean stratification, particularly if the equator experiences cooling. Conversely, warming by increasing stellar irradiation strongly warms equatorial waters while deep water formed at high latitudes remains near the freezing point. This leads to enhanced stratification over most of the ocean. 

\subsection{Obliquity} \label{sec:obliquity}
Increasing obliquity from 0-45$^{\circ}$ yields warmer climates and a reduction of sea ice on annual average (\citealt{kang_mechanisms_2019}; Figure \ref{fig:obliq}f), both of which contribute to enhanced gas exchange kinetics (Figure \ref{fig:obliq}f). The equator-to-pole temperature difference is also substantially reduced due to a more equal distribution of stellar irradiation at the planet's surface, leading to reduced ocean stratification (Figure \ref{fig:obliq}e). In our highest obliquity scenario, the summer pole becomes warmer than the equator. Nonetheless, the mixed layer depth decreases on long-term and global average with increasing obliquity. This is somewhat unexpected given the dramatic reduction in density stratification (Figure \ref{fig:obliq}e), but may be partially explained by a reduction in wind stress as obliquity increases (Figure \ref{fig:obliq}d). Moreover, the depth of the mixed layer is strongly seasonal, deepening by as much as a factor of 100x in the winter compared to the warm summer in our 45$^{\circ}$ obliquity scenario (Figure \ref{fig:seasons}). 

Globally upwelling increases only slightly with increasing obliquity (Figure \ref{fig:obliq}a). In our highest obliquity scenario the spatial distribution of upwelling varies seasonally due to changes in surface currents (Figure \ref{fig:o45_circ}). These patterns may allow seasonal nutrient supply over large regions of the ocean compared to low-obliquity scenarios. Moreover, extreme seasonal deepening of the mixed layer may allow entrainment of nutrients from depth independent of upwelling and may provide a key mechanism for nutrient regeneration on high obliquity planets. 

\subsection{Salinity} \label{sec:salinity}
\begin{figure*}[t]
\centering
\includegraphics[width=\textwidth]{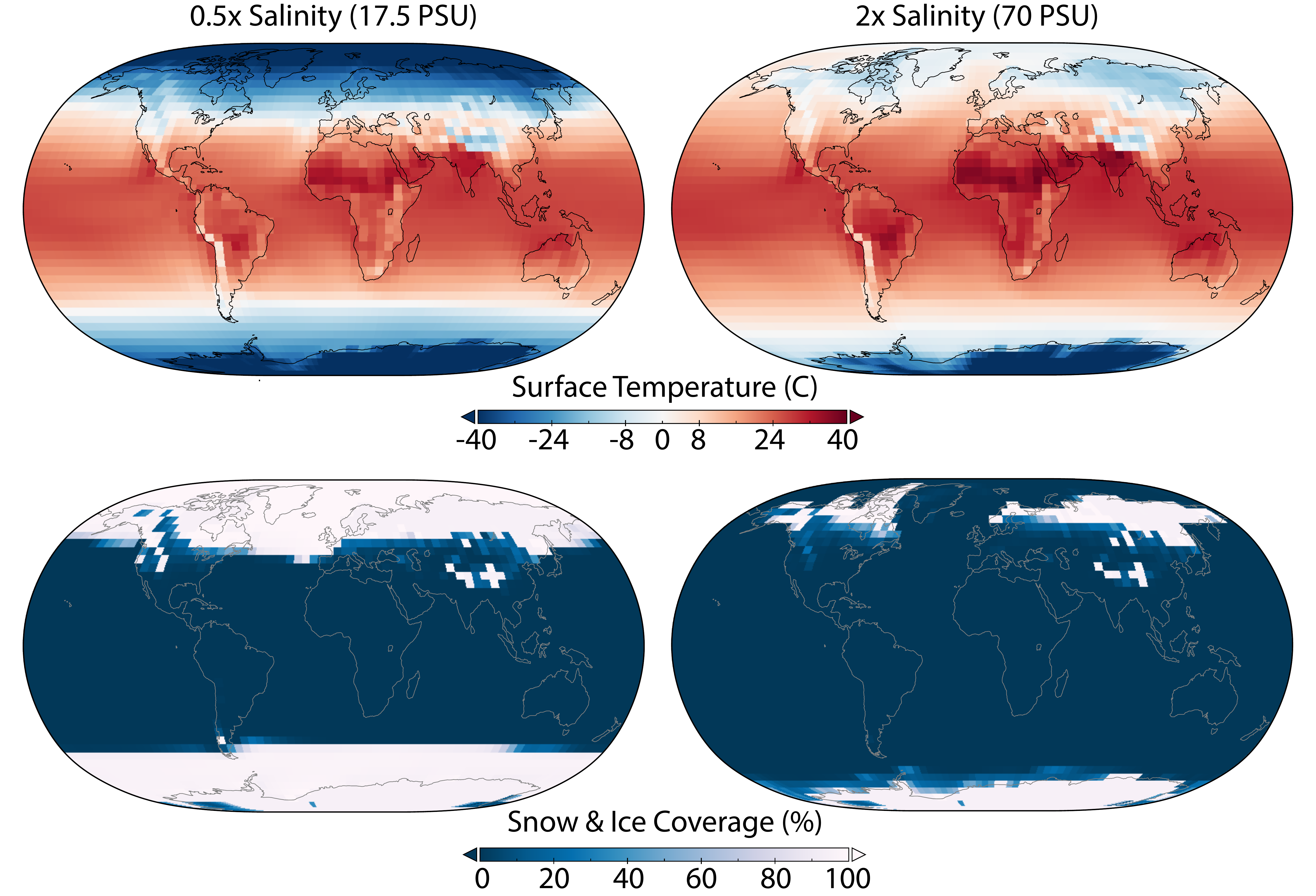}
\caption{\textbf{Comparison of surface air temperature (top) and snow/ice cover (bottom) for low and high salinity scenarios}. Increasing ocean salinity results in warming and limits sea ice formation.}
\label{fig:salt_comp}
\end{figure*}

Ocean salinity impacts the climate system in several ways. For example, salt strongly influences temperature--density relationships and the density structure of the ocean. However, the most significant impact that varying salinity has on the marine environment in our experiments is its influence on sea ice formation: relatively small increases in salinity result in dramatic reductions in sea ice (Figure \ref{fig:salt}f. There are two reasons. First, salt suppresses the freezing point of seawater and thus directly limits sea ice formation. Moreover, exclusion of salt during sea ice formation (`brine rejection') produces high density water that sinks at high latitudes. Brine rejection may trigger deep convection locally, bringing up relatively warm water from below. Enhanced sinking at high latitudes also strengthens the global overturning circulation, increasing upwelling at low latitudes and driving the flow of warm surface water poleward \citep{cael_oceans_2017}. Each of these effects interact with the ice-albedo feedback, which tends to amplify changes in ice coverage through associated changes in planetary albedo. Global-average temperature ultimately increases with ocean salinity because the reduction of ice coverage results in a less reflective surface and higher water vapor content of the atmosphere. Doubling ocean salinity compared to present-day Earth yields 6 K warming on global average and precludes sea ice formation (Figure \ref{fig:salt}f). This warming is strongest in the Arctic but extends into the mid and low latitudes (Figure \ref{fig:salt_comp}).     

The combination of the inhibition of freezing and a larger thermal expansion coefficient at higher salinities allow for a larger temperature and density contrast laterally within the surface ocean, ultimately enhancing vertical density stratification throughout the ocean (Figure \ref{fig:salt}e). The result is a shallowing of the mixed layer depth with increasing ocean salinity (Figure \ref{fig:salt}b). An accompanying reduction of the atmospheric equator-to-pole temperature gradient and weakened wind stress reinforces this effect (Figure \ref{fig:salt}d; Table  \ref{table:atmdata}). 

Upwelling at the base of the mixed layer increases slightly with increasing salinity despite decreasing wind stress (Figure \ref{fig:salt}a). This increase likely reflects an increase in the brine-driven circulation discussed above. Indeed, gains in upwelling diminish at high salinity as sea ice formation wanes and brine rejection ceases. The oceanographic and climatic consequences of salinity on planets lacking sea ice warrants further investigation. 

Sea-air gas exchange is enhanced with increasing salinity due to warmer temperatures on global average and reduced sea ice cover (Figure \ref{fig:salt}c,f). Decreasing gas solubility with increasing salinity would also favor more efficient transfer of biological gases to the atmosphere from saltier oceans, but is not accounted for in our $k_{O_{2}}$ metric.  

\section{Discussion}
\subsection{Oceanographic Constraints on Life} \label{sec:bio_disc}
We simulated a diversity of habitable ocean environments, some of which may be more hospitable to large, productive biospheres than others. In particular, we hypothesize that planets with more efficient nutrient recycling via ocean upwelling will be better hosts for photosynthetic life than planets where nutrients will be sequestered at depth. Our results suggest that slowly rotating planets with higher surface pressure may support the most active biospheres because ocean upwelling---and thus nutrient recycling---is maximized under these conditions. Upwelling is also enhanced on planets somewhat larger than Earth, in salty oceans, and at intermediate positions within the Habitable Zone. 

Higher orbital obliquity also promotes nutrient return from depth via greater upwelling with foci that shift seasonally. Even more intriguingly, large seasonal differences in the mixed layer on high obliquity planets may allow entrainment of previously exported nutrients into the mixed layer. This seasonality may also provide an efficient mechanism for the transfer of biosignatures produced in the deep ocean such as CH$_4$ to the atmosphere, particularly given that high obliquity disfavors sea ice. High obliquity planets may thus host particularly active, globally distributed life that may be uniquely detectable \citep{olson_atmospheric_2018}, but strong seasonality in light availability may have additional biological consequences that are not considered here. An additional caveat may be that on very high (\textgreater45$^{\circ}$) obliquity planets, stratospheric wettening may enhance water loss and limit the duration of planetary habitability \citep{kang_wetter_2019}.  

Although we did not explicitly vary land area or continental distribution, our results highlight the importance of continents for habitability (e.g., \citealt{ward_rare_2000}). Coastal upwelling was the largest contributor to global upwelling in all of our model scenarios. Moreover, in addition to promoting coastal upwelling and the recycling of nutrients from depth, continental weathering plays a key role in nutrient delivery to the ocean to balance nutrient burial in marine sediments on geological timescales. This is in addition to the key role that continental weathering plays in climate regulation \citep{abbot_indication_2012}. Although ocean worlds may meet existing definitions of habitability in some circumstances \citep{kite_habitability_2018}, such planets will not be favorable targets for life detection owing to inevitable limitations on biospheric productivity in the absence of continents and associated nutrient fluxes via upwelling and weathering. Future work should explore the sensitivity of upwelling to alternative continental configurations and whether there exists optimal or problematic land distributions \citep{lingam_dependence_2019}. 

A related issue is ocean depth. We did not explore the consequences of changing water inventories, but dramatically differing ocean volumes are likely as a result of stochastic water delivery \citep{ramirez_habitable_2014, luger_extreme_2015, tian_water_2015}, and/or surface-mantle exchange \citep{cowan_water_2014, schaefer_persistence_2015, komacek_effect_2016}. Ocean depth may affect the interplay between the biological pump and ocean circulation. A very shallow ocean that permits benthic photosynthesis and/or wind-mixing of the entire water column may optimize photosynthetic rates by minimizing nutrient export to dark depths. However, net O$_2$ production requires spatial separation of photosynthetic O$_2$ and biomass. While shallow oceans may be good for gross biospheric productivity, it is net export production that ultimately favors remote detectability (e.g., via biogenic chemical disequilibrium in the atmosphere; \citep{krissansen-totton_disequilibrium_2018}). Export and associated biosignature accumulation may therefore be limited in very shallow oceans. Conversely, a very deep ocean provides a large reservoir for nutrient accumulation while also limiting nutrient fluxes from the weathering of exposed continental crust \citep{kite_habitability_2018, lingam_dependence_2019}. In the extreme case of a planet with 100 Earth oceans of water, stabilization of high density ices at the bottom of the ocean may further limit nutrient supply by inhibiting water-rock interactions \citep{kitzmann_unstable_2015}. Unfortunately, constraining exoplanet ocean depth may not be feasible with foreseeable instrumentation \citep{kite_habitability_2018} beyond the context provided by the inferred presence/absence of exposed continents from observations \citep{deeg_mapping_2018}. Further work is necessary to illuminate the fate of nutrients in very shallow or deep oceans, determine the potential impact of exo-ocean depth on planetary suitability for large photosynthetic biospheres, and understand the uncertainty that unknown ocean depth contributes to habitability and biosignature characterization. 

We also did not consider the impact of synchronous rotation and/or changes to the stellar spectrum. Extrapolating the results of our study, synchronously rotating planets may favor greater ocean upwelling. Additionally, strong tidal mixing on these worlds may enhance nutrient recycling \citep{lingam_implications_2018}, providing attractive habitats for photosynthetic life. However, life on synchronously rotating planets may experience light (rather than nutrient) limitation \citep{lehmer_productivity_2018}. It is thus unclear how our results might be extended to M star systems. Future work should include explicit representation of light and nutrient limited photosynthesis and elucidate the planetary circumstances for which each ingredient is likely to be limiting for photosynthesis globally and on long-term average.

Finally, one important caveat for extending our results to exoplanet life detection is that the conditions that favor the maintenance of globally productive biospheres, such as nutrient recycling, may differ from the conditions that favor the origin of life. A planet must meet both criteria to host remotely detectable life, but we do not know how/where life originated on Earth or which conditions are most suitable for the origin of life on other planets. Our study thus focuses on the surface conditions that may be most conducive to the global success of Earth-like photosynthetic life that may be uniquely detectable \citep{schwieterman_exoplanet_2018}---but these worlds may differ from those on which life is most likely to originate in the first place. These distinctions are not reflected in the prevailing binary view of habitability based on the stability of liquid water; moving forward it will be important to distinguish between planets that are conducive to the emergence of life vs. those that are survivable by life (i.e., conventional HZ planets) vs. those that may be particularly hospitable for life (`superhabitable' planets; \cite{heller_superhabitable_2014, del_genio_climates_2019}) vs. those that are uniquely favorable for \textit{detectable} life as discussed here. 

\subsection{Observational Opportunities and Challenges} \label{sec:obs_disc}
Some of the planetary parameters that favor ocean upwelling and productive marine biospheres will be remotely observable. For example, Rayleigh scattering, pressure broadening of absorption bands, and pressure-sensitive dimers may reveal the presence of a dense atmosphere \citep{misra_using_2014}. In particular, although N$_2$ itself is not spectrally active, the N$_2$-N$_2$ collisional pair is spectrally recognizable and may be used to constrain N$_2$ levels \citep{schwieterman_detecting_2015}. Although high surface pressure is not a guarantee of a hospitable marine environment, detection of a Rayleigh slope or N$_2$-N$_2$ absorption would demonstrate the existence of an atmosphere and may be an indication of surface conditions that promote wind-driven upwelling, nutrient recycling in the ocean, and biospheric productivity. Constraints on surface pressure therefore may be a useful consideration before dedicating limited observing time to the search for biosignatures such as seasonality that require long integration times. 

Time-resolved observations provide an opportunity to simultaneously probe rotation rate and to assess continentality by enabling longitudinal mapping of ocean vs. land \citep{cowan_alien_2009, deeg_mapping_2018, lustig-yaeger_detecting_2018}. Recall that long day-length and continentality both favor nutrient recycling via upwelling. Additionally, slower rotation and the presence of continents both enhance meridional heat transport and reduce ice coverage. In combination, observational constraints on rotation rate and the presence of continents may suggest the potential for active, globally distributed life in an ocean that is not too deep and may communicate effectively with the overlying atmosphere--all of which would limit the possibility of a biosignature false negative. 

Unfortunately, not all planetary parameters that we explored will be readily observable. Exo-ocean salinity will likely be impossible to constrain observationally, but observations that indicate liquid water on a very cold planet may be suggestive of high salinity. The salinity of Earth's ocean has changed dramatically in our history (e.g., \citealt{yang_persistence_2017}) and the salinities of other oceans in our own solar system apparently vary widely (e.g, \citealt{hand_empirical_2007, postberg_salt-water_2011, mitri_shape_2014}). Critically, we lack predictive models for these differences. Uncertainties regarding exo-ocean salinity must be considered in future attempts to simulate the climates of potentially habitable exoplanets and to delineate the boundaries of the Habitable Zone \citep{cullum_importance_2016,cael_oceans_2017,del_genio_habitable_2019}. 

\subsection{Implications for Earth history} 
In addition to informing exoplanet characterization and life detection efforts, our study may provide insight to Earth's history. Throughout its roughly 4 billion years of inhabitation, we know that Earth's rotation rate has slowed, its surface pressure has fluctuated, the salinity of its ocean has varied, stellar irradiation has steadily increased, and continental distributions have continuously evolved. These histories imply that ocean circulation patterns, including upwelling, may have varied dramatically in our planet's past. These changes come with biogeochemical impacts.  

In particular, evolving ocean circulation may have consequences for Earth's oxygenation. Several lines of evidence point to an origin of oxygenic photosynthesis very early in Earth's history, potentially up to half a billion years before low-level oxygenation of the atmosphere during to the Great Oxidation Event \citep{planavsky_evidence_2014}, but the reason photosynthesis failed to oxygenate Earth's atmosphere for so long is not understood. The reasons why post-GOE oxygen stabilized at levels much lower than today are even more enigmatic \citep{planavsky_low_2014}, but emerging models apparently require that primary productivity was lower than today for much of Earth's history (e.g., \citealt{ozaki_sluggish_2019}). Low nutrient levels are thus widely invoked to explain limited surface oxygenation despite oxygen production \citep{reinhard_evolution_2017, laakso_limitations_2018, ozaki_sluggish_2019, guilbaud_phosphorus-limited_2020}, but debate remains regarding the physical mechanism for limiting nutrient supply. Our results may provide an intriguing path forward: ocean upwelling and associated nutrient recycling processes may have simply been less efficient on an early Earth that rotated faster \citep{williams_geological_2000, bartlett_analysis_2016}, had lower surface pressure \citep{som_earths_2016, lehmer_atmospheric_2020, payne_oxidized_2020}, had less continental exposure \citep{johnson_limited_2020}, and orbited a fainter star compared to present day Earth \citep{gough_solar_1981}. The steady slowing of Earth's rotation, the growth of the continents, and a continuously brightening Sun may have manifested as a secular increase in nutrient recycling, stimulating photosynthesis and promoting the long-term oxygenation of the atmosphere. 

\section{Conclusions} \label{sec:concl} 
Ocean circulation controls the distribution and activity of life on Earth, and it modulates the communication between life in the ocean and the overlying atmosphere. Ocean circulation ultimately throttles the accumulation of biological products in planetary atmospheres and is thus an important consideration for the oxygenation of our planet and the detectability of exoplanet life. We used an ocean-atmosphere GCM to explore ocean dynamics and the resulting ocean habitats on planets differing from Earth. Our analysis focused on three ocean characteristics of biogeochemical significance, including: gas exchange kinetics, mixed layer depth, and upwelling at the base of the mixed layer. An intriguing result of our modeling is that the most Earth-like scenario was sub-optimal for nutrient recycling and biosignature transfer to the atmosphere in many of our sensitivity experiments, introducing the possibility that true Earth twins may not be the most favorable targets for exoplanet life detection missions. Ocean circulation patterns on planets that rotate more slowly, have higher surface pressure, higher orbital obliquity, and saltier oceans than Earth may be more conducive to nutrient regeneration, biospheric productivity, and atmospheric biosignature accumulation than our own planet. Planets with larger radii may also be appealing candidates. Moving forward, we must make a distinction between worlds that meet some minimum criteria to be considered habitable (e.g., possessing liquid water) and those that will be most hospitable to globally productive, \textit{remotely detectable} life. Oceanographic phenomena should be at the center of such efforts.

\subsection*{Acknowledgements}
S.L.O acknowledges support from the T.C. Chamberlin Postdoctoral Fellowship in the Department of the Geophysical Sciences at the University of Chicago. We thank the University of Chicago Research Computing Center for providing computing resources that were essential to this work. This work was partially supported by the NASA Astrobiology Program Grant Number 80NSSC18K0829 and benefited from participation in the NASA Nexus for Exoplanet Systems Science research coordination network. 

\bibliography{Exo-oceanography}
\end{document}